# Recent Advances in Mobile Grid and Cloud Computing

Sayed Chhattan Shah
Department of Information Communications Engineering
Hankuk University of Foreign Studies, South Korea
shah@hufs.ac.kr

**Abstract**

Grid and cloud computing systems have been extensively used to solve large and complex problems in science and engineering areas. These systems include powerful computing resources connected through high speed networks. Due to recent advances in mobile computing and networking technologies it has become feasible to integrate various mobile devices such as robots, aerial vehicles, sensors, and smart phones with grid and cloud computing systems. This integration enables design and development of next generation of applications through sharing of resources in mobile environments and also introduces several challenges due to dynamic and unpredictable network. This paper discusses applications, research challenges involved in design and development of mobile grid and cloud computing systems, and recent advances in the field.

## 1. Introduction

A distributed system consists of a collection of autonomous computers, connected through a network and distribution middleware, which enables computers to coordinate their activities and to share the resources of the system, so that users perceive the system as a single, integrated computing facility.

One of the best examples of distributed system is an automated teller machine where user is provided with a simple and easy to use interface to perform numerous transactions. From user's point view there is a single system but in background hundreds of different computing devices are connected through various networks to provide a range of financial services.

The distributed systems are divided into three main categories: cluster, grid and cloud computing system. In cluster, distributed computing devices are connected through a high-speed local area network whereas in grid, geographically distributed resources are connected through a wide area network to solve large and complex problems.

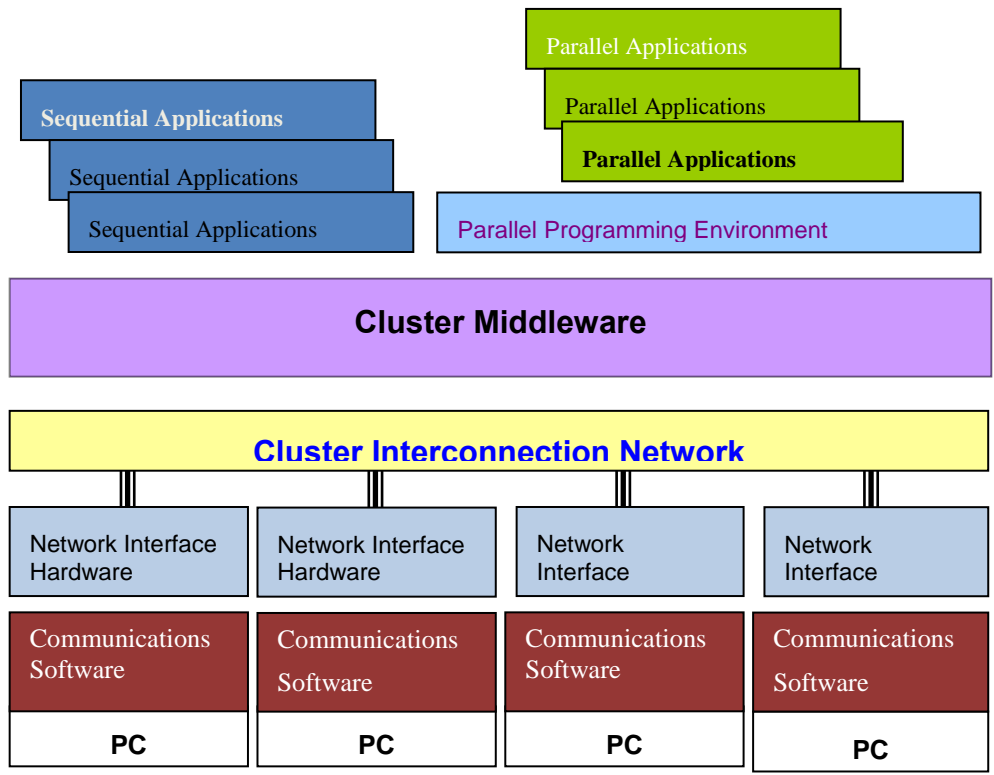

Figure 1: Cluster Architecture

A layered architecture of cluster and grid computing systems is presented in Figures 1 and 2, respectively. The resource layer includes computing devices which are connected through networking and communication technologies. The middleware layer provides resource and task management services which include resource monitoring and discovery service, fault management service, resource allocation service and task migration service. In addition, middleware hides all the complexities and provides a single system image to user and applications running on the system. Resource and task management in clusters is usually based on a centralized architecture while grid relies on distributed architectures for resource and task management.

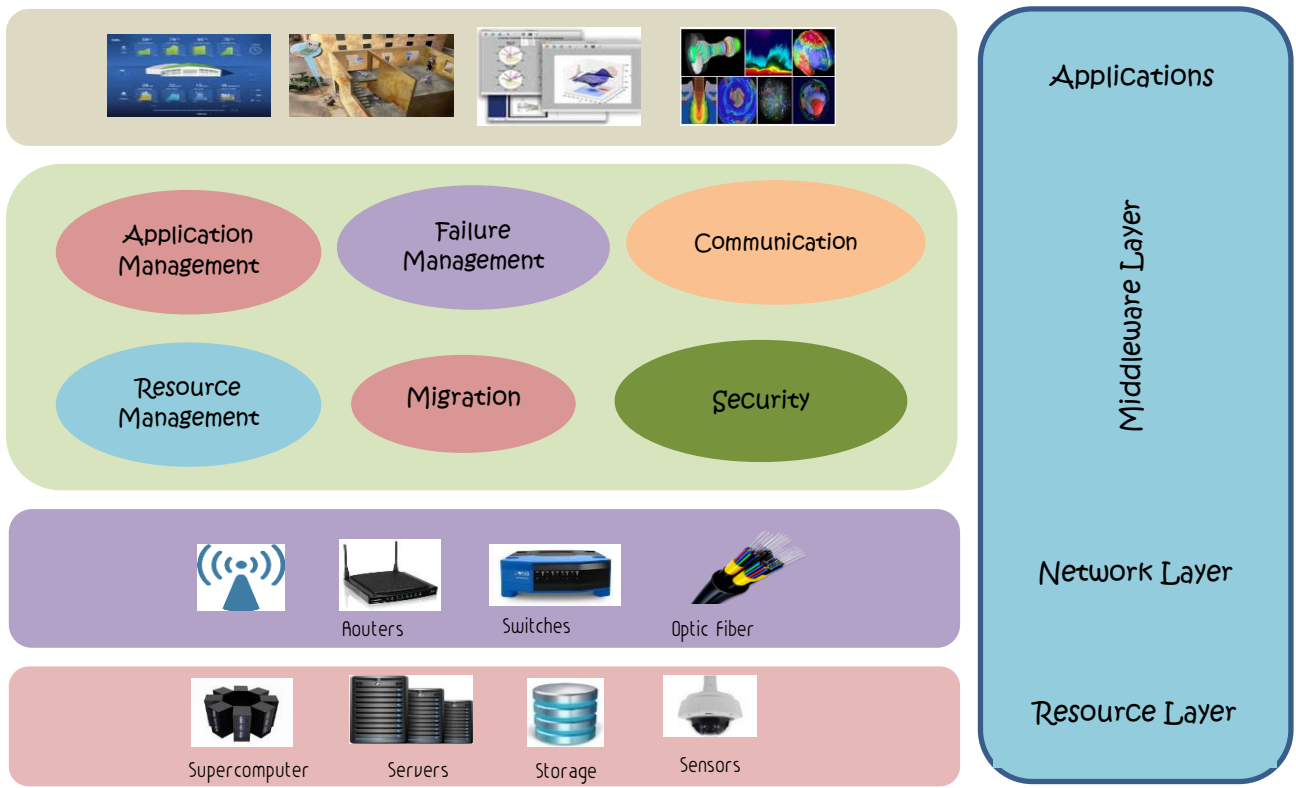

Figure 2: Grid Architecture

The cloud computing has evolved from cluster and grid computing and is integration of various concepts and technologies such as hardware virtualization, utility computing, autonomic computing, pervasive computing and service-oriented architecture. In cloud computing everything from computing power to communication infrastructure and applications are delivered as a service over a network. Cloud computing technologies and service model are given in Figure 3.

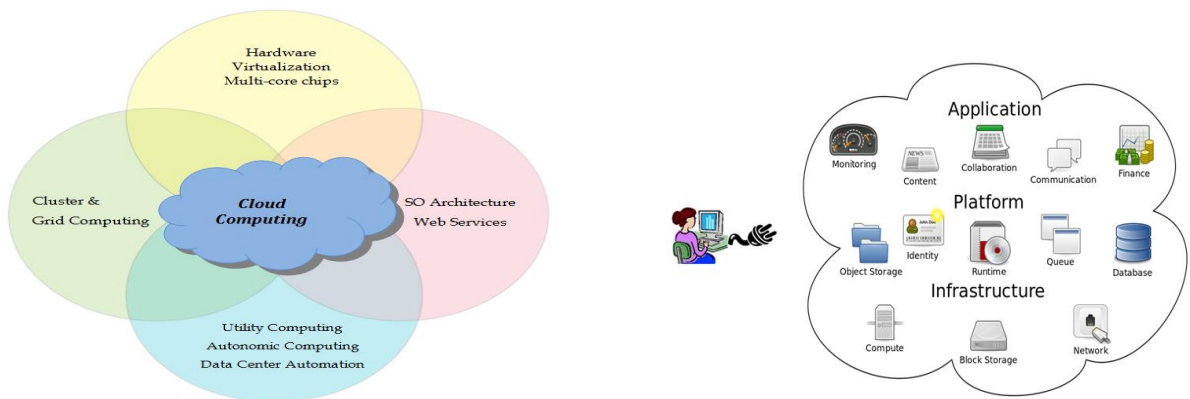

Figure 3: Cloud Technologies and Service Model

The basic idea is to provide computing and communication resources on demand for a fee just like the electrical power grid. User should be provided with an easy to use interface to access vast amount of cloud resources without being concerned about details such as how power is generated or how various resources are connected.

Grid and cloud computing systems have been extensively used to solve large and complex problems in science and engineering areas such as drug design, earthquake simulation, and climate modeling. These systems include powerful computing resources connected through high speed networks. Applications of grid and cloud computing systems in science and engineering areas are presented in Figure 4.

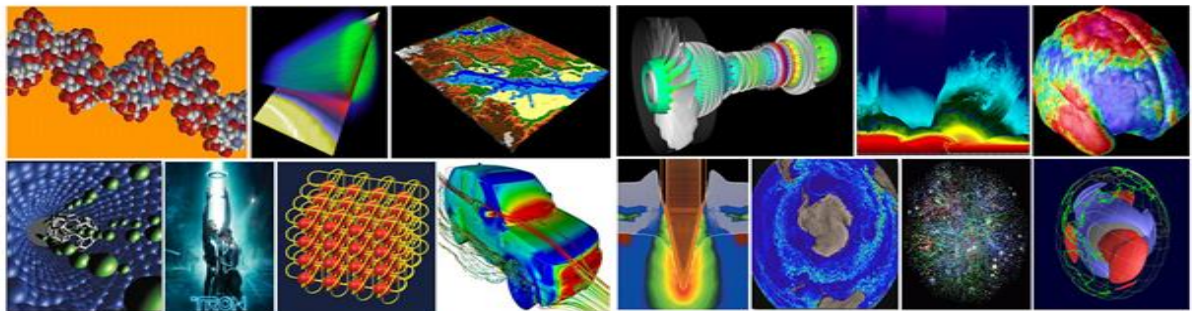

Figure 4: Illustrate applications of grid and cloud computing systems in science and engineering areas

Due to recent advances in mobile computing and networking technologies it has become feasible to integrate various mobile devices such as robots, aerial vehicles, sensors, and smart phones with grid and cloud computing systems. The approaches for integrating mobile devices with grid and cloud computing systems are divided into two main categories: Mobile grid and cloud computing and mobile ad hoc grid and cloud computing. Both categories are further divided into two sub-categories: data grid and cloud, and computational grid and cloud. Taxonomy of distributed systems is given in Figure 5.

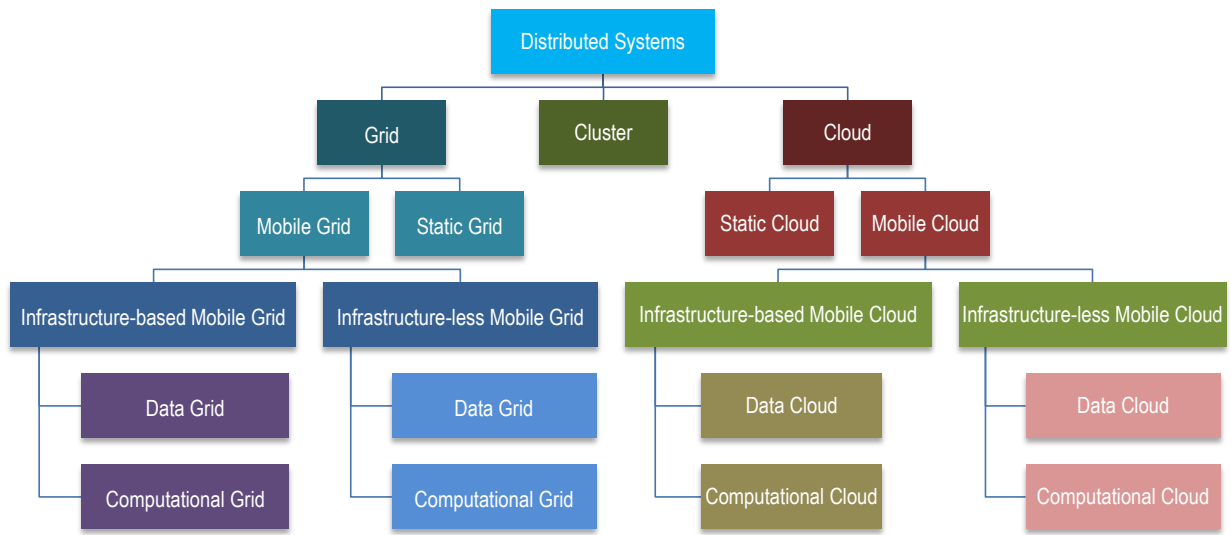

Figure 5: Taxonomy of Distributed Systems

## 2. Mobile Grid and Cloud Computing

*Mobile Cloud Computing*

In mobile cloud computing, mobile devices are integrated with a cloud computing system through an infrastructure-based communication network such as cellular network. The computationally intensive tasks are offloaded to a cloud for execution. Architecture of mobile cloud system is given in Figure 6.

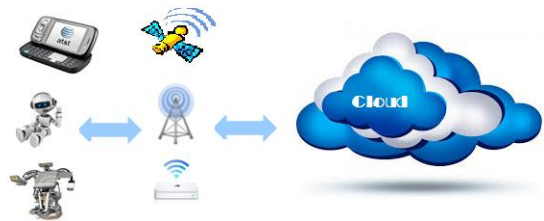

Figure 6: Mobile Cloud System Architecture

*Mobile Grid and Cluster Computing*

In mobile grid computing, mobile devices are connected to a grid computing system through an infrastructure-based communication network such as cellular network. The computationally intensive tasks are sent to grid computing system which after processing sends results back to the mobile device. In mobile cluster computing, mobile devices are connected to cluster through an infrastructure-based network. Architecture of mobile grid system is given in Figure 7.

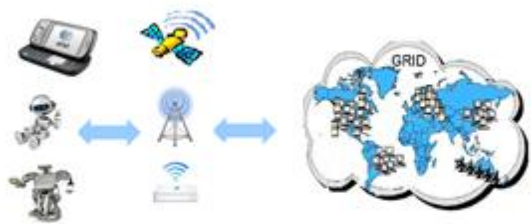

The continuous increase in processing power of mobile devices, improved data transfer rates of wireless networks, and growing popularity of cloud computing systems and mobile devices has made it feasible and valuable to integrate mobile devices with traditional distributed systems. The currently available mobile device such as iPhone 6 has more processing power than the supercomputer used in the early nineties [1] [2] while the current generation of wireless network such as 4G provides 100 megabits per second of data rate which is enough for streaming a video. Several companies are already testing 5G networks that aim to provide 10 gigabits per second of data rate. Figure 8 illustrates advances in mobile computing and communication technologies, worldwide forecast of mobile phones, and cloud computing market share.

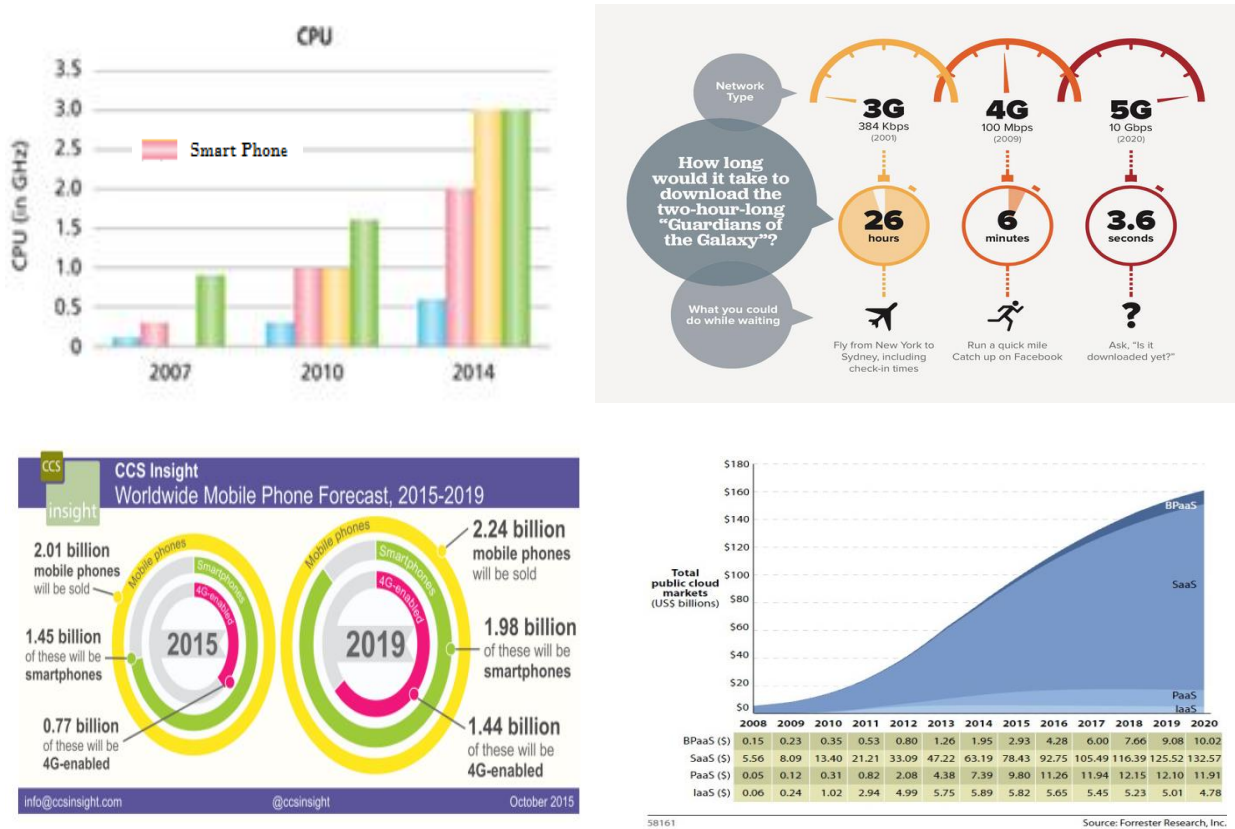

Figure 8: Illustrates advances in mobile computing and communication technologies, worldwide forecast of mobile phones, and cloud computing market share.

*Benefits*

Cloud computing systems provide vast amount of storage space and processing power on demand. Integration of mobile devices with cloud would enable mobile devices to access vast data storage capacity and processing power. In addition, mobile user would be able to execute data and computationally intensive applications such as image and video processing on mobile devices. Data storage and execution of such applications on cloud would also improve reliability and would extend the battery life of mobile devices.

*Cloud Robotics*

Cloud robotics is another interesting topic in area of mobile cloud computing. In cloud robotics, robots are integrated with cloud computing system to access vast amount of processing power and data storage. This integration allows robots to offload heavy tasks such as image processing and voice recognition to cloud. Block diagram of cloud robotics is given in Figure 9.

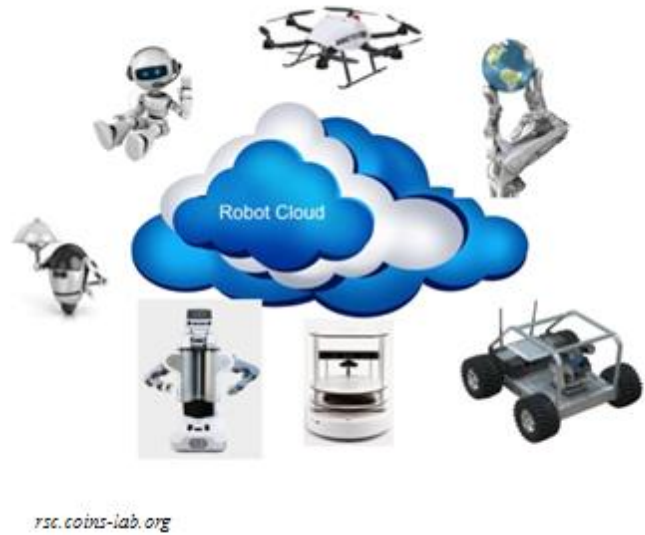

Figure 9: Cloud Robotics

In addition to advantages discussed for mobile cloud computing, integration of robots with cloud provides several other advantages.

1- Execution of computationally intensive tasks on cloud would result in cheaper, lighter and easy-to-maintain hardware, longer battery life, minimum software updates, and hassle-free and invisible CPU hardware upgrades.

2- Shared object library: Assume a scenario in which a robot deployed in an urban environment encounters an unknown object and therefore is unable to manipulate it. To address this issue one option is to maintain information of commonly used objects on a robot which would require large amount of storage space and also processing power. The second option is to maintain the information of commonly used objects on the cloud. When a robot encounters an unknown object, it can take a picture of object and send it to cloud which in turn will return the required information such as object name, 3D model, mass, materials, friction properties, and usage instructions required to operate the object. The object library on the cloud can be shared by several robots and can be regularly updated by experts.

3- Like object library, a library of numerous skills, navigation algorithms, and maps can also be maintained on the cloud. When a robot is deployed in unknown environment and is unable to navigate, it can directly download the required map and navigation algorithm to navigate in the environment. In addition, library may include several navigation algorithms developed by numerous experts. For example, when robot has a limited battery power it may choose to download an energy efficient algorithm and if energy is not a concern a computationally intensive algorithm with a better performance can be used.

*Sensor Grid and Cloud*

In sensor grid and cloud, sensor nodes are integrated with grid and cloud computing system through a gateway node. Sensor nodes are used to collect data which is then sent to cloud for further processing. Sensor cloud architecture is presented in Figure 10. Sensor cloud have several applications in numerous areas including cyber physical systems such as smart home, smart grid and smart city [28]. For example, in smart aging project several bio and environmental sensors such as camera, noise detector, temperature monitor, heart rate and blood pressure monitors are deployed to observe home environment and physiological health of an individual. The data collected by sensors is sent to an application on cloud where numerous algorithms for emotions and sentiments detection, activity recognition, and situation detection are applied to provide healthcare and emergency services and to manage resources at the home.

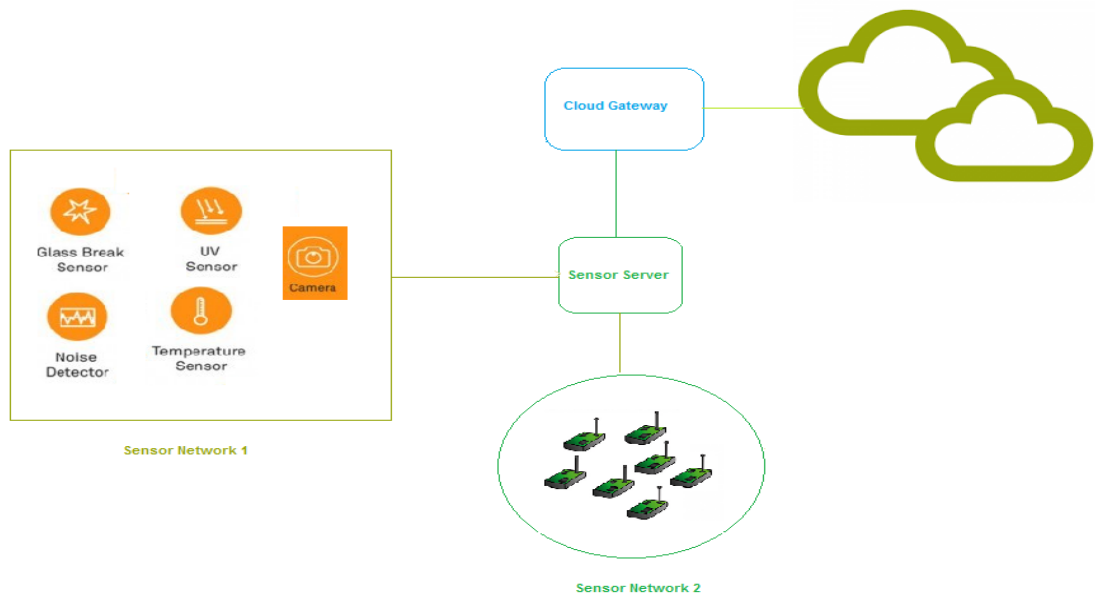

Figure 10: Sensor Cloud Architecture

## 3. Mobile Ad hoc Computational Grid and Cloud

The mobile grid and cloud computing systems are restricted to infrastructure-based communication systems such as cellular network, and therefore cannot be used in mobile ad hoc environments. A mobile ad hoc computational grid [29] is a distributed computing infrastructure that enables mobile nodes to share computing resources without pre-existing network infrastructure whereas mobile ad hoc computational cloud is a distributed computing infrastructure in which multiple mobile devices interconnected through a mobile ad hoc network are presented *as one powerful, unified computing resource*. The system architecture of mobile ad hoc computational cloud is given in Figure 11.

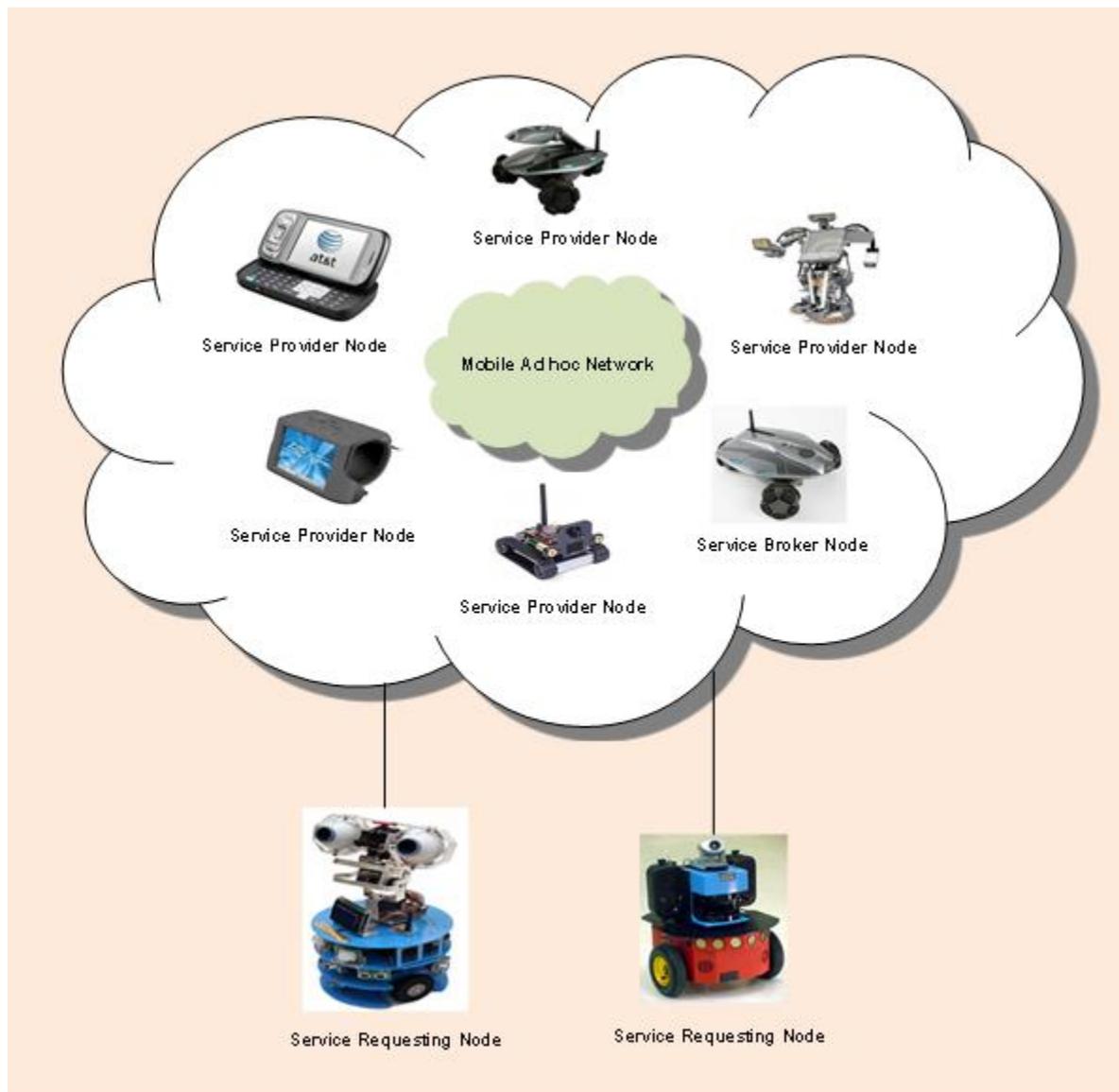

Figure 11: illustrates a diagram of mobile ad hoc computational cloud in which multiple mobile devices interconnected through a mobile ad hoc network are presented *as one powerful, unified computing resource*.

Mobile ad hoc computational grid is a combination of computational grid and mobile ad hoc network. Computational grid is a software infrastructure that allows distributed computing devices to share computing resources to solve computationally-intensive problems. Whereas mobile ad hoc network is a wireless network of mobile devices that communicate with each other without any pre-existing network infrastructure.

Architecture of mobile ad hoc computational grid is presented in Figure 12. Mobile nodes communicate with each other through a mobile ad hoc network which provides several communication services including transport, routing, and medium access. The middleware layer, computational grid, provides resource management service which includes resource monitoring, resource discovery, and task management services, fault management service, mobility management service, communication service, and task migration service. In addition, middleware hides all the complexities and provides a single system image to user and applications running on the system.

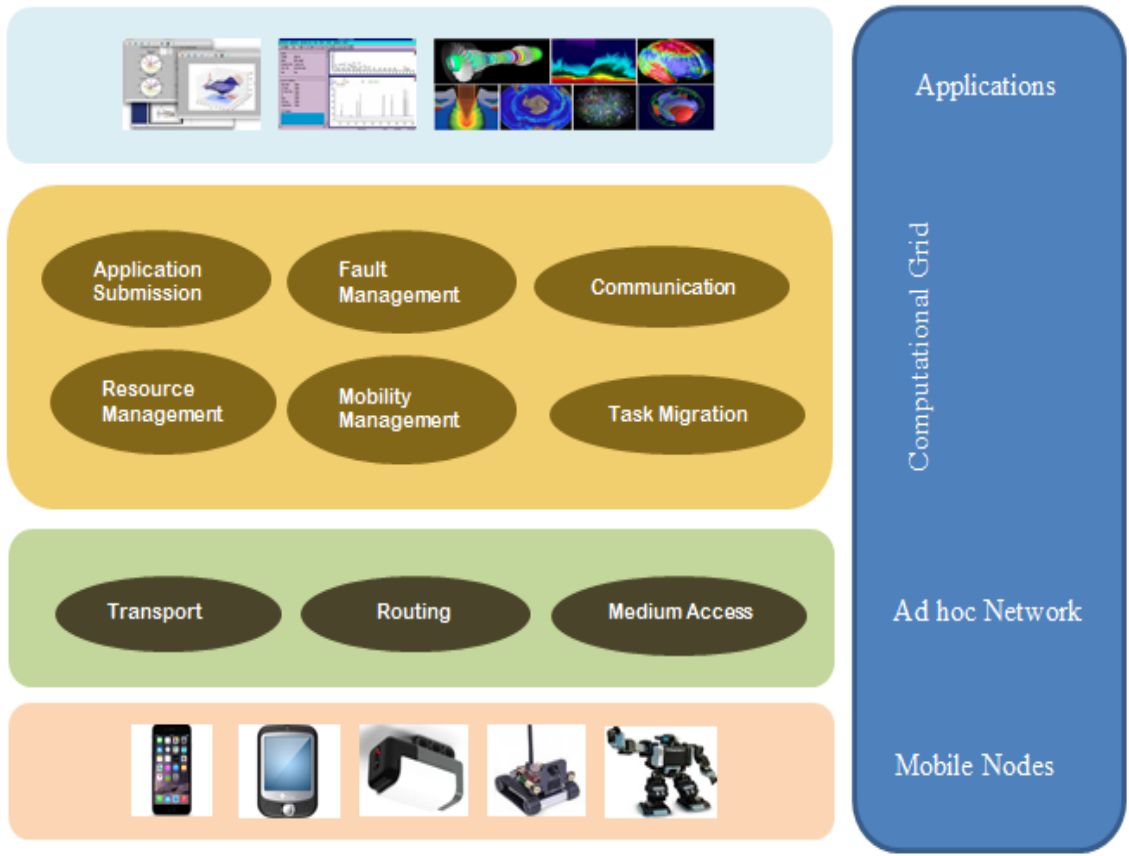

Figure 12: Mobile Ad hoc Computational Grid Architecture

*Applications*

(a) *Mobile Automated Intelligent Video Surveillance System*

Success of a military or disaster relief operation depends on several factors including understanding of the physical environment and real-time detection and tracking of mobile and stationary targets. In order to understand the environment, various mobile robots and micro drones equipped with audio, video and environmental sensors are deployed to collect data which is then processed to construct a three dimensional map of environment in real-time. The collected data is also used to detect and track mobile and stationary targets which involve sophisticated image and video processing algorithms [5-7].

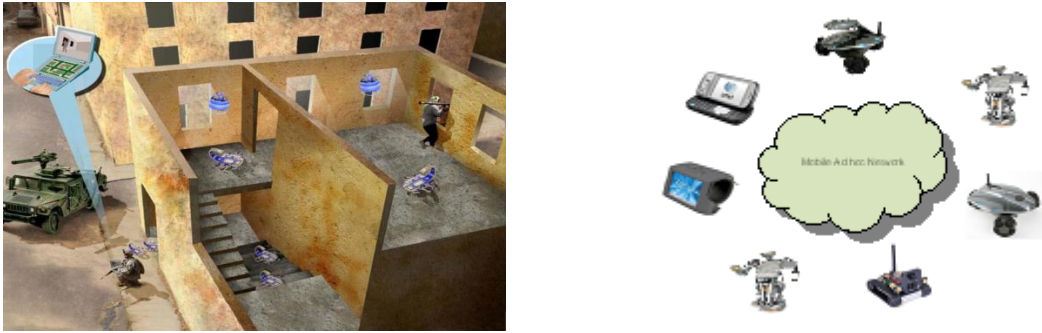

Figure 13: Illustrates Automated Intelligent Video Surveillance System and Mobile Ad hoc Grid Architecture

In order to perform these tasks a vast amount of computing and storage resources are required. To address the issue there are two options: 1- Send data to an application on cloud through infrastructure based communication system such as cellular network. 2 - Create a virtual supercomputing node comprised of mobile robots, micro drones, and soldiers' wearable devices.

The first option is not feasible because in battlefield or disaster relief operation, infrastructure-based communication systems are usually not available and even if they are available their communication performance is much lower then short range wireless communication technologies such as Wi-Fi Direct. For example, compared to fourth generation networks that provide 100 megabits per second of data rate, wireless local area network provides up to 600 megabits per second of data rate. Figure 13 illustrates automated intelligent video surveillance system and mobile ad hoc grid architecture.

*(b) Mobile Gaming System*

A group of friends are on a trip and would like to play a game in which human body tracking is used to animate characters. To track human body, mobile devices are equipped with video sensors and we assume that one mobile device has a projector for display. This scenario requires tracking of human body parts and other game related tasks in real-time. In order to realize this application a group of mobile nodes can share their computing resources on demand to form a grid. From grid, a sub-group of mobile devices can be assigned to each user to track his body movements and at the same time these devices can be used to execute other game related tasks and communicate actions.

## 4. Research Challenges

Compared to traditional parallel and distributed computing systems such as grid and cloud, mobile grid and cloud computing systems introduce several challenges due to global and local node mobility, high latency, limited power, and dynamic network environment. This section discusses the key research challenges.

- **Global and local node mobility:** Global node mobility refers to the movement of nodes across the coverage area whereas local node mobility refers to the movement of nodes within the coverage area. Global node mobility results in failure of one or several tasks whereas local node mobility may increase data transfer cost and thus task completion time. Effect of local node mobility is described in Figure 14. Task migration or reallocation strategy can be used to manage task failure but this will introduce delay and will increase task completion time.

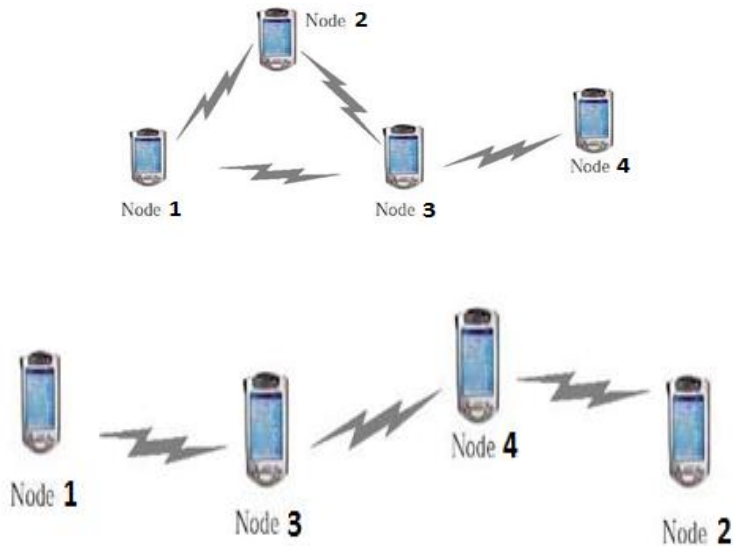

Figure 14: illustrates effect of local node mobility. Tasks executing on Node 1 and Node 2 are communicating directly. During execution of task, Node 2 moves within the network coverage area and therefore is not directly accessible to Node 1. So, task executing on Node 1 communicates with a task executing on Node 2 through two intermediate nodes, Node 3 and 4. This would increase data transfer time and thus task completion time.

Global node mobility also introduces the problem of outdated information managed by resource discovery and monitoring system. To address the problem, continuous monitoring of resources is required but it will increase communication cost. An alternate approach is to enquire the status of resources during decision making but this will introduce delay.

To address the problems, an effective and robust resource allocation scheme is proposed in [23]. The scheme uses history of user's mobility patterns to select nodes that will remain connected for long period of time. Since this scheme relies on user's mobility history, it cannot be used in situations where mobility history is not available or user has random and irregular mobility patterns.

- **Dynamic communication environment:** Communication system of mobile ad hoc grid and cloud is unpredictable, dynamic and limited in performance due to low bandwidth, local and global node mobility, and lack of preexisting network infrastructure. The connection quality at different network sections fluctuates over the time. Even different nodes may experience different connection quality at the same time. Due to these reasons, data transfer cost is critical for application performance.

- **Power management:** Communication energy consumption depends on two key factors: transmission power required to transmit data and communication cost induced by data transfers between tasks. In literature, several mechanisms are proposed to address the problem but most of them are focused on the conservation of processing energy, while saving communication energy remains an open problem which becomes even more critical for data-oriented applications.

  To reduce communication energy consumption, an energy efficient resource allocation scheme is developed in [22]. The main idea is to use transmission power control mechanism and allocate dependent tasks to nodes that are accessible at minimum transmission power. However, further investigation is required to incorporate the transmission power control mechanism in resource discovery, resource monitoring, failure management, and task migration services.

- **Task migration:** Task migration primarily involves transferring both code and data across the nodes. Tasks are migrated for several reasons such as to improve application performance and resource utilization, balance load, conserve energy, and manage task failure. Compared to traditional systems, design of migration service for mobile grid and cloud computing systems is difficult due to dynamic network environment. The most common migration strategy is to estimate migration cost and determine task execution time with or without the migration. The estimation of migration cost depends on estimation of data transfer time which is not straightforward particularly for data-oriented applications.

- **Computation offloading:** One of the key challenges in mobile grid and cloud computing systems is to devise an efficient task offloading strategy. In literature, several offloading strategies [39-49] are proposed which are classified based numerous factors such as:

  o *Offloading objective:* An application *[39-41]* or part of application *[42-49]* is offloaded to a remote system for several reasons such as to improve application performance, reduce energy consumption, or to avoid task failure.

- *Offloading client:* Mobile node that uses offloading could be smart phone, mobile robot, autonomous aerial vehicle, or sensor.

- *Infrastructure for offloading:* Mobile node may offload computation to a cluster [31-32, 51-52], grid [36-38], cloud computing system [39-42], or to a powerful computer connected through a wide or local area network [40].

- *Type of applications:* Various application types have been studied such as computationally intensive applications, data intensive applications, data parallel applications, and task parallel applications [49].

- *Application partitioning mechanisms:* To partition an application into tasks two key approaches have been proposed: static and dynamic. The later one is more flexible and suitable for dynamic and unpredictable environments.

- *Offloading decision:* A decision to offload task could be made before runtime or during runtime.

In order to design an effective offloading strategy, several factors such as network bandwidth, available energy, processing speed of remote and host resource, task queue size, amount of data transfers between a task on client node and a task to be allocated on remote server node should be considered.

- **Architecture:** Every architecture have some pros and cons, for example, centralized architecture results in effective resource allocation decisions due to network wide view but suffers from scalability and single point of failure problems. Whereas distributed architecture solves scalability and single point of failure problems but results into poor resource allocation decisions due to lack of network wide view. Possible architectures for mobile ad hoc grid and cloud computing system are given in Figure 15. There is a need to develop an energy efficient, effective and robust architecture that should give maximum performance in a wide range of scenarios.

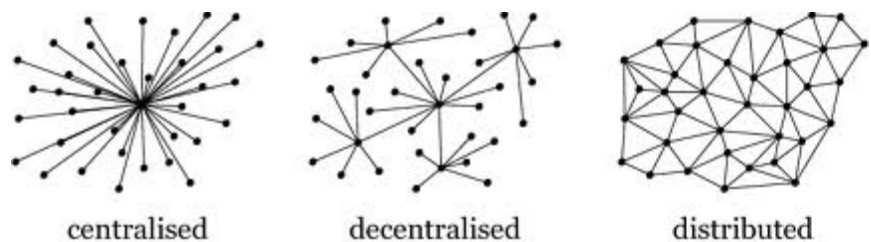

Figure 15: Architectural choices for mobile ad hoc grid and cloud computing system

- **Parallel programming model:** Programming models developed for infrastructure-based computing systems cannot be used in dynamic and ad hoc network environments. There is need to design a programming model which provide lightweight migration and communication mechanisms and should support high mobility scenarios.

- **Transport and routing protocols:** Numerous transport and routing protocols have been developed either to reduce data transfer times or transmission energy consumption. Most of these protocols are not adaptive to application or system's requirements. There is need to develop adaptive transport and routing protocols. Such protocols, for example, should reduce energy consumption either using multi hop communication or transmission power control mechanism if that does not cause tasks to miss the deadline, or send data through multiple routes when network becomes unstable. In addition, the protocols should actively communicate with grid and mobility management systems in order to provide reliable and robust services.

- **Wireless communication technologies:** Several wireless communication technologies such as Wi-Fi Direct and Bluetooth with different characteristics are available to transfer data from one node to another in an ad hoc mode. In addition, most of the devices provide an option to use and switch between multiple connections at the same time. In order to provide efficient and reliable data transmission, further work is required to investigate and adopt wireless communication technologies for mobile ad hoc grid and cloud.

- **Development of simulation environment:** Existing simulators allow simulation of either networking protocols or cloud related algorithms and schemes. Simulators that support simulation of both, networking protocols and cloud management systems, are not designed for mobile ad hoc environments. Further work is required for the development of environment in which both networking and computational cloud related algorithms can be evaluated.

## 5. Selected Research Projects

*Cloud Robotics*

Researchers at Social Robotics Lab have built a cloud computing infrastructure to generate 3-D models of environments allowing robots to perform simultaneous localization and mapping much faster than by relying on their onboard computers. SLAM refers to a technique for a robot to build a map of the environment without a priori knowledge, and to simultaneously localize itself in the unknown environment.

At CNRS, researchers are creating object databases for robots to simplify the planning of manipulation tasks like opening a door. The idea is to develop a software framework where objects come with a "user manual" for the robot to manipulate them.

Gostai, a French robotics firm, has built a cloud robotics infrastructure called GostaiNet, which allows a robot to perform complex tasks such as speech recognition, advanced vision and face detection remotely on a cloud. Architecture of GostaiNet system is presented in Figure 16.

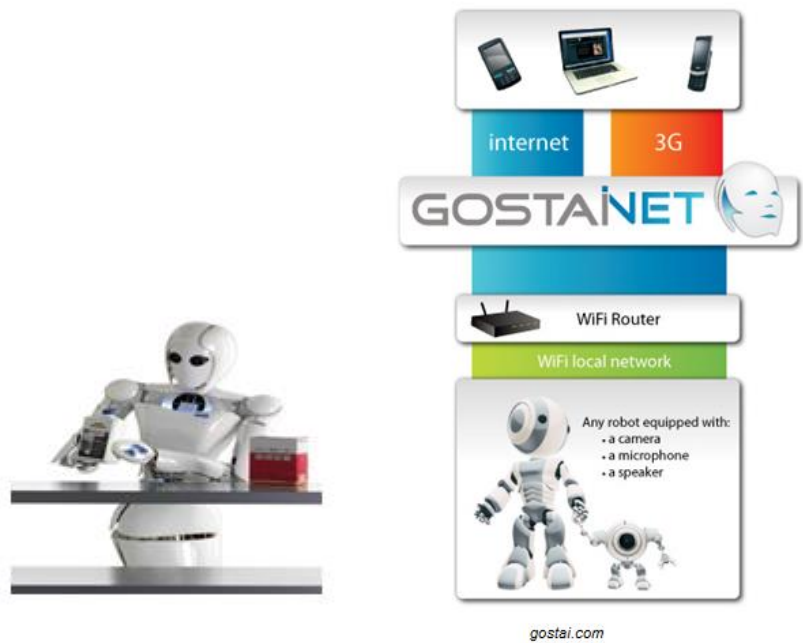

Figure 16: GostaiNet System Architecture

*Mobile Grid and Cloud Computing*

Collaborative Drones

The objective of this project was to develop an aerial surveillance system that can be used in disaster management situation or military operation. The video data collected by unmanned aerial vehicles is submitted to a ground system that processes the data to detect and track objects such as cars or persons in real-time [27].

SINUS

Self-organizing Intelligent Network of Autonomous Unmanned Aerial Vehicles [25] is another project that aims to provide: 1) reliable aerial networking for robust multimedia streaming, 2) distributed coordination of unmanned aerial vehicles movement and task execution, and 3) system integration.

Fare-Share

Researchers at Cyber Physical Systems Laboratory at Rutgers University are working on a project that aims to exploit collective capabilities of nearby mobile and stationary devices to execute computationally intensive models for deriving physiological parameters and for acquiring context awareness in real time [24]. Block diagram of Fair- Share project is given in Figure 17.

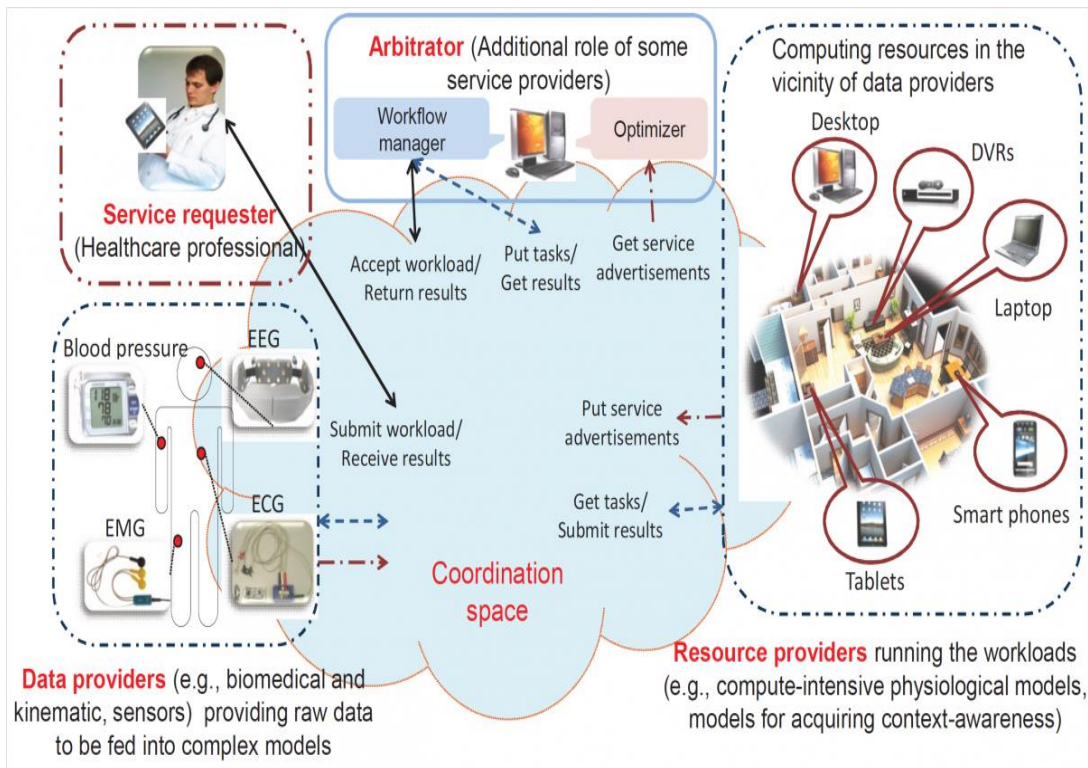

Figure 17: Block Diagram of Fair- Share Project

Content Based Mobile Edge Networking

In traditional systems, soldiers have to visit battlefield camp to obtain latest battlefield contents because information required at the tactical edge is not immediately available. CBMEN program aims to develop technologies to allow rapid sharing of up-to-date imagery, maps and other vital information directly among front-line units. The mobile device of a solider will generate, distribute and maintain contents at the tactical edge [26].

## 6. Literature Review

*Mobile Grid and Cloud Computing*

In literature, several architectures have been proposed to enable mobile devices to share resources either with pre-existing network infrastructure-based computing systems [30-33, 41-48] such as cloud or with other mobile devices using short range wireless communication technologies [22-23, 29, 34-39].

One such architecture is proposed in [20] in which mobile devices are connected to a locally available powerful computer named cloudlet through short range wireless communication technologies such as Wi-Fi. Cloudlet is then connected to a cloud computing system through a satellite. The aim is to reduce satellite communication latency. Cloudlet system architecture is presented in Figure 18.

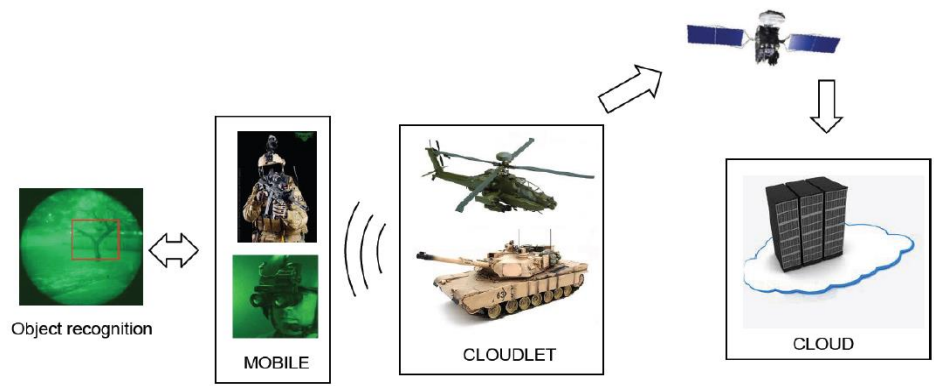

Figure 18: Cloudlet System Architecture

Object recognition application has been implemented to evaluate the performance. The solider captures and sends image of an individual to cloudlet which preprocesses the image and sends it to cloud computing system through a satellite for further processing.

In [50] authors have proposed a system that allows students to offload image and video processing tasks on internet cloud. In [51] a single hop ad hoc cloud has been suggested to compute a climate model. In proposed scenario, train commuters form an ad hoc cloud comprised of smart phones which is used to forecast local weather and ozone concentrations at the destination using data collected and transmitted by passengers in another train coming from the destination. A hybrid cluster comprised of stationary and mobile nodes interconnected by wireless and wired networks has been proposed in [52]. The proposed system could be used by an architect equipped with wearable computing devices to execute a computationally intensive 3D rendering application. Six online and batch scheduling heuristics are proposed in [53] to offload computationally intensive independent tasks on mobile ad hoc cloud. MinHop allocates task to a node accessible through minimum number of hops. Minimum Execution Time with Communication algorithm selects a node which takes least amount of time to execute a task whereas Minimum Completion Time with Communication algorithm assigns task to a node with minimum expected completion time. Both online scheduling algorithms take into account computation and communication cost but former focuses on execution time while later one on completion time. For batch scheduling, MinMinComm and MaxMinComm algorithms are proposed. MinMinComm estimates completion time of each task on each node and selects the node with minimum earliest completion time while MaxMinComm focuses on maximum earliest completion time. The proposed scheduling heuristics are based on traditional MET, MCT, MinMin, and MaxMin heuristics and are extended to consider communication cost when estimating task completion time. The problem of deciding whether to execute tasks locally or on a remote system has been investigated in [54]. The evaluations suggest that tasks with high computation-to-communication ratio should be executed on a powerful remote system. The scheme proposed in [55] performs task allocations under varying assumptions about the connectivity environment. For example, in the ideal network environment where the future contact can be accurately predicted, the authors proposed a greedy task allocation algorithm that iteratively chooses the destination node for every task with the minimum task completion time. The scheme is based on a decentralized architecture and it does not provide any mechanism to address node mobility. In [56], various job stealing techniques such as random stealing, best rank aware stealing and worst rank aware stealing based on a centralized architecture were developed to reduce the processing energy consumption. The rank is calculated based on a benchmark factor for each device, the estimated uptime of a device and the number of jobs assigned to that device. In [57] three offloading architecture for mobile clouds are described: (i) offloading to a cloud server accessible through a wide area network, (ii) offloading to a nearby server accessible through a local area network, and (iii) offloading to nearby mobile devices. The problem of unpredictable network connectivity, node mobility, energy consumption, and device failure has been addressed in [58].

*Sensor Cloud*

Several approaches have been proposed to integrate sensing nodes with grid and cloud computing systems. For example, a hybrid static and mobile grid computing system is proposed in [16] in which mobile and static computing devices and bio-sensing nodes are integrated and presented as a one unified system. The bio-sensors collect vital signs such as blood pressure, temperature, electrocardiogram, and oxygen saturation of an individual. The collected data is processed and analyzed on mobile grid computing infrastructure in order to determine the health of an individual. To deal with uncertainty, an idea of application waypoints has been introduced in which service provider executing application task reports to the broker with an estimate of residual task completion time. If the broker does not receive feedback about the estimated residual task completion time from the service provides at the specified waypoint, it marks service provider as failed and assigns additional resources to take over the incomplete tasks. A resource allocation algorithm to efficiently process telemedicine data in the grid is proposed in [21]. In proposed algorithm sensors attached to patient's body collect and send health related data to grid through a mobile device. A patient management application deployed on the grid processes and analysis the patient's data.

A sensor-cloud infrastructure proposed in [15] integrates sensors with cloud computing system. In sensor-cloud infrastructure physical sensors integrated with cloud computing system are virtualized as virtual sensors and are provided as a service. A pull-based resource allocation algorithm is proposed in [17] in which a service provider node pulls tasks from the service broker nodes, executes them and submits results once task completes its execution. In [18] authors have proposed a sensor grid platform to combine real-time data about the environment with vast computational resources. The proposed sensor grid platform can be deployed using centralized architecture or decentralized architecture. In centralized architecture, a sensor network connected to grid collects data while processing of data is carried out on the grid. In distributed architecture, a sensor network collects data and performs simple data processing tasks while computationally intensive data processing and analysis tasks are executed on the grid.

A sensor data collection network to integrate sensor data into grid applications is discussed in [19]. The sensor data collection network includes three key components: the data collection network, sensor entry points, and application entry points. The data collection network discovers, filters, and queries multiple sensor networks. Each sensor network has one or more sensor entry points that map application data requirements onto low-level sensor network operations. The application entry points provide application connectivity to data collection network. The proposed system is based on publish-subscribe paradigm. A sensor network publishes sensor data and metadata through service entry points while application subscribes to sensor network and receives a data in real time. In [20] authors have proposed a scalable proxy-based architecture for sensor grid in which nodes in wireless sensor network are provided as a service on the grid. The use of proxy-based architecture where proxy acts as an interface between grid and sensor network supports a wide range of sensor network implementations. All these systems and schemes including [4-14] assume pre-existing network infrastructure and therefore are not suitable for ad hoc network environments.

## 7. Conclusion

Due to recent advances in mobile computing and communication technologies it has become feasible to develop next generation of distributed computing systems where mobile nodes such as robots and aerial vehicles should be able to share resources either with pre-existing network infrastructure-based computing systems such as cloud or with other mobile nodes using short range wireless communication technologies such as Wi-Fi Direct.

Compared to traditional parallel and distributed computing systems, mobile grid and cloud computing systems pose numerous challenges due to node mobility, shared and unreliable communication medium, low bandwidth, high latency, limited battery power, and infrastructure-less network environment. This paper discussed applications of mobile grid and cloud computing systems, research challenges, future research directions, and recent advances in the field.

**Acknowledgment**

This work was supported by Hankuk University of Foreign Studies Research Fund of 2016.

**Conflict of Interests Disclosure**

The author declares that there is no conflict of interest regarding the publication of this paper.

**References**

[1] A modern smartphone or a vintage supercomputer, http://www.phonearena.com/news/A-modern-smartphone-or-a-vintage-supercomputer-which-is-more-powerful_id57149, Accessed in Jan 2016.

[2] A supercomputer in your pocket, http://www.charliewhite.net/2013/09/smartphones-vs-supercomputers/, Accessed in Jan 2016.

[3] How 5G will push a supercharged network to your phone, home, car, http://www.cnet.com/news/how-5g-will-push-a-supercharged-network-to-your-phone-home-and-car/, Accessed in Jan 2016.

[4] B. G. Chun and P. Maniatis, Augmented Smartphone Applications through Clone Cloud Execution, 12[th] Workshop on Hot Topics in Operating Systems, Switzerland: USENIX, 2009

[5] S. Garriss, R. Caceres, S. Berger, R. Sailer, L. van Doorn, and ˆX. Zhang, Trustworthy and Personalized Computing on Public Kiosks, 6[th] International Conference on Mobile Systems, Applications, and Services (MobiSys '08), 2008, pp. 199 – 210

[6] M. Satyanarayanan, P. Bahl, R. Caceres, N. Davies, the Case for VM-Based Cloudlets in Mobile Computing, IEEE Pervasive Computing, 8 (4) 14–23, 2009

[7] J. Rellermeyer, O. Riva, and G. Alonso, AlfredO: Architecture for Flexible Interaction with Electronic Devices, 9[th] ACM/IFIP/USENIX International Conference on Middleware (Middleware 2008), Lecture Notes in Computer Science, Vol. 5346. Leuven, Belgium: Springer, 2008, pp. 22–41.

[8] J. S. Rellermeyer, G. Alonso, and T. Roscoe, R-OSGi: Distributed Applications through Software Modularization, ACM/IFIP/USENIX 8[th] International Middleware Conference (Middleware 2007). Newport Beach, CA, USA: Springer, Nov. 2007, pp. 50–54.

[9] R. K. Balan, M. Satyanarayanan, S. Y. Park, and T. Okoshi, Tactics-based Remote Execution for Mobile Computing, Int. Conf. Mobile Systems, Applications and Services, San Francisco, May 5–8, 2003.

[10] B.-G. Chun, S. Ihm, P. Maniatis, M. Naik, A. Patti. CloneCloud: elastic execution between mobile device and cloud. ACM EuroSys, 2011

[11] E. Cuervo, A. Balasubramanian, D. K. Cho, A. Wolman, S. Saroiu, R. Chandra, and P. Bahl. Maui: making smartphone last longer with code offload. In ACM MobiSys, 2010

[12] M.A.M. Mohamed et al., MOSET: An anonymous remote mobile cluster computing paradigm, Journal of Parallel Distributed Computing, 65 (2005) 1212 – 1222

[13] Z.-L. Zong, M. Nijim, and X. Qin, Energy-Efficient Scheduling for Parallel Applications on Mobile Clusters, Cluster Computing: The Journal of Networks, Software Tools and Applications, 11 (1), 91 - 113, 2008.

[14] Chunlin Li; Layuan Li, Tradeoffs between energy consumption and QoS in mobile grid, Journal of Supercomputing, 55 (3), 2011.

[15] M. Yuriyama and T. Kushida, "Sensor-cloud infrastructure: physical- sensor management with virtualized sensors on cloud computing," in Network-Based Information Systems (NBiS), 13th International Conference on. IEEE, 2010, pp. 1-8

[16] H. Viswanathan, E. K. Lee, and D. Pompili, "Mobile grid computing for data- and patient-centric ubiquitous healthcare," in Proc. of IEEE Workshop on Enabling Technologies for Smartphone and Internet of Things, Seoul, Korea, June 2012.

[17] H. Kim, Y. el Khamra, I. Rodero, S. Jha, and M. Parashar, "Autonomic Management of Application Workflows on Hybrid Computing Infrastructure," Telecomm. Sys., vol. 19, no. 2-3, pp. 75–89, Feb. 2011.

[18] C. K. Tham and R. Buyya (2005), SensorGrid: Integrating Sensor Networks and Grid Computing. CSI Communications, 29(1):24-29, July 2005.

[19] M. Gaynor, S.L. Moulton, M. Welsh, E. LaCombe, A. Rowan, J. Wynne, Integrating wireless sensor networks with the grid, IEEE Internet Computing, 8 (4) (2004), pp. 32–39

[20] H.B. Lim, P. Mukherjee1, V.T Lam, W.F. Wong and S. See., Sensor Grid: Integration of Wireless Sensor Networks and the Grid, Proceedings of 30th IEEE Conference on Local Computer Networks, pp. 91-98, Sydney, Australia, November 2005.

[21] T. Vigneswari and M. A. Maluk Mohamed, "Scheduling in Sensor Grid Middleware for Telemedicine Using ABC Algorithm," International Journal of Telemedicine and Applications, Vol. 2014, 7 pages, 2014

[22] Sayed Chhattan Shah, Myong-Soon Park, " An Energy-Efficient Resource Allocation Scheme for Mobile Ad Hoc Computational Grids", Journal of Grid Computing Journal no.10723 (SCI-E), Apr.16,2011

[23] S C Shah et al., An effective and robust two-phase resource allocation scheme for interdependent tasks in mobile ad hoc computational Grids, Journal of Parallel and Distributed Computing, 2012

[24] Uncertainty-aware Resource Provisioning in Mobile Computing Grids for Real-time In-situ Data Processing, http://nsfcac.rutgers.edu/cps/projects/uncertainty-aware-resource-provisioning-mobile-computing-grids-real-time-situ-data, Accessed on Jan 2016.

[25] SINUS, http://uav.lakeside-labs.com/overview/sinus/, Accessed in Jan 2016.

[26] Content-Based Mobile Edge Networking Program, http://www.darpa.mil/program/content-based-mobile-edge-networking

[27] Collaborative Drones, http://uav.lakeside-labs.com/overview/cdrones/, Accessed in Jan 2016.

[28] iCore cognitive framework, http://www.iot-icore.eu, Accessed in Jan 2016.

[29] Sayed Chhattan Shah, Energy Efficient and Robust Allocation of Interdependent Tasks on Mobile Ad hoc Computational Grid, Concurrency and Computation: Practice and Experience, 27(5) 1226-1254, 2015.

[30] B.-G. Chun and P. Maniatis, "Augmented Smartphone Applications Through Clone Cloud Execution," in Proceedings of the 12th Workshop on Hot Topics in Operating Systems, USENIX, 2009.

[31] H. Zheng, R. Buyya, S. Bhattacharya, "Mobile cluster computing and timeliness issues", Informatica: International J. Comput. Inform. 23 (1)

[32] M.A.M. Mohamed et al., "MOSET: An anonymous remote mobile cluster computing paradigm", Journal of Parallel Distributed Computing, 65 (2005) 1212 – 1222

[33] Z.-L. Zong, M. Nijim, and X. Qin, "Energy-Efficient Scheduling for Parallel Applications on Mobile Clusters", Cluster Computing: The Journal of Networks, Software Tools and Applications, vol. 11, no. 1, pp. 91 - 113, March 2008

[34] K. A. Hummel and G. Jelleschitz., Robust De-centralized Job Scheduling Approach for Mobile Peers in Ad Hoc Grids, 7[th] IEEE Int. Symp. Cluster Computing and the Grid, May 14–17, 2007.

[35] V. Vetri Selvi, S. Sharfraz, and R. Parthasarathi, "Mobile Ad Hoc Grid Using Trace Based Mobility Model," GPC 2007, LNCS 4459, pp. 274–285, 2007

[36] Piotr Grabowski, Bartosz Lewandowski Poznan "Access from J2me-Enabled Mobile Devices to Grid Services" Supercomputing and Networking Center, Noskowskiego Poznan, 61-704 Poland

[37] Fox, G. Ho, A. Rui Wang Chu, E. Isaac Kwan, A collaborative sensor grids framework, IEEE CTS 2008

[38] Jamie M. Robinson et al, "Sensor Networks and Grid Middleware for Laboratory Monitoring", First International Conference on e-Science and Grid Computing (e-Science'05), 2005

[39] S. Shilve, H.J. Siegel, A.A. Maciejewski, P. Sugavanam, T. Banka, R. Castain, K. Chindam, S. Dussinger, P. Pichumani, P. Satyasekaran, W. Saylor, D. Sendek, J. Sousa, J. Sridharan, and J. Velazco, "Static Allocation of Resources to Communicating Subtasks in a Heterogeneous Ad Hoc Grid Environment," Journal of Parallel Distributed Computing, vol. 66, no. 4, pp. 600–611, 2006.

[40] S. Garriss, R. Caceres, S. Berger, R. Sailer, L. van Doorn, and ´X. Zhang, "Trustworthy and Personalized Computing on Public Kiosks," in Proceeding of the 6th International Conference on Mobile Systems, Applications, and Services (MobiSys '08). Breckenridge, CO, USA: ACM, 2008, pp. 199 – 210

[41] M. Satyanarayanan, P. Bahl, R. Caceres, and N. Davies, "The Case for VM-Based Cloudlets in Mobile Computing," IEEE Pervasive Computing, vol. 8, no. 4, pp. 14–23, Oct. 2009

[42] I. Giurgiu, O. Riva, D. Juric, I. Krivulev, and G. Alonso, "Calling the Cloud: Enabling Mobile Phones as Interfaces to Cloud Applications," in Proceedings of the 10th ACM/IFIP/USENIX International Conference on Middleware (Middleware '09). Urbana Champaign, IL, USA: Springer, Nov. 2009, pp. 1–20.

[43] J. Rellermeyer, O. Riva, and G. Alonso, "AlfredO: An Architecture for Flexible Interaction with Electronic Devices," in Proceedings of 9th ACM/IFIP/USENIX International Conference on Middleware (Middleware 2008), ser. Lecture Notes in Computer Science, vol. 5346.Leuven, Belgium: Springer, 2008, pp. 22–41.

[44] J. S. Rellermeyer, M. Duller, and G. Alonso, "Engineering the Cloud from Software Modules," in Proceedings of the Workshop on Software Engineering Challenges in Cloud Computing (ICSE-Cloud, in conjunction with ICSE 2009). Vancouver, Canada: IEEE, 2009, pp. 32–37.

[45] J. S. Rellermeyer, G. Alonso, and T. Roscoe, "R-OSGi: Distributed Applications through Software Modularization," in Proceedings of the ACM/IFIP/USENIX 8th International Middleware Conference (Middleware 2007). Newport Beach, CA, USA: Springer, Nov. 2007, pp. 50–54.

[46] R. K. Balan, M. Satyanarayanan, S. Y. Park, and T. Okoshi, "Tactics-based Remote Execution for Mobile Computing," 1st Int. Conf. Mobile Systems, Applications and Services, San Francisco, May 5–8, 2003.

[47] B.-G. Chun, S. Ihm, P. Maniatis, M. Naik, A. Patti. CloneCloud: elastic execution between mobile device and cloud. In ACM EuroSys, 2011

[48] E. Cuervo, A. Balasubramanian, Cho, A. Wolman, S. Saroiu, R. Chandra, P. Bahl. Maui: making smartphone last longer with code offload. In ACM MobiSys, 2010

[49] H. T. Dinh, C. Lee, D. Niyato, and P. Wang, "A survey of mobile cloud computing: Architecture, applications, and approaches," Wireless Communications and Mobile Computing (WCMC), accepted

[50] Ferzli, R.; Khalife, I., "Mobile cloud computing educational tool for image/video processing algorithms," in *Digital Signal Processing Workshop and IEEE Signal Processing Education Workshop (DSP/SPE), 2011 IEEE* , vol., no., pp.529-533, 4-7 Jan. 2011

[51] Busching, F.; Schildt, S.; Wolf, L., "DroidCluster: Towards Smartphone Cluster Computing - The Streets are Paved with Potential Computer Clusters," in *Distributed Computing Systems Workshops (ICDCSW), 2012 32nd International Conference on*, vol., no., pp.114-117, 18-21 June 2012

[52] Cheng, L.; Wanchoo, A.; Marsic, I., "Hybrid cluster computing with mobile objects," *Proceeding of 4[th] International Conference on High Performance Computing in Asia-Pacific Region*, vol.2, 909-914, 14-17 May 2000

[53] Bo Li, Yijian Pei, Hao Wu, Bin Shen., Heuristics to allocate high-performance cloudlets for computation offloading in mobile ad hoc clouds, The Journal of Supercomputing, 71(8), pp 3009-3036 2015.

[54] Kumar, Karthik, and Yung-Hsiang Lu, Cloud computing for mobile users: Can offloading computation save energy? Computer 4 (2010): 51-56.

[55] Cong Shi, Vasileios Lakafosis, Mostafa H. Ammar, Ellen W. Zegura, Serendipity: enabling remote computing among intermittently connected mobile devices, Proceedings of the thirteenth ACM international symposium on Mobile Ad Hoc Networking and Computing, 2012

[56] J.M. Rodriguez, C. Mateos and A. Zunino, Energy-efficient Job Stealing for CPU-intensive processing in Mobile Devices, Computing, Springer, In Press, 2012.

[57] Simin Ghasemi-Falavarjani, Mohammad Ali, and Behrouz Shahgholi, Context-aware multi-objective resource allocation in mobile cloud, Computer and Electrical Engineering, 44 (2015) 218-240

[58] Hariharasudhan Viswanathan, Eun Kyung Lee, Ivan Rodero, and Dario Pompili, Uncertainty-Aware Autonomic Resource Provisioning for Mobile Cloud Computing, IEEE Transactions on Parallel and Distributed Systems, 26(8), August 2015.